\documentclass[12pt]{article}
\usepackage{latexsym}
\usepackage[english]{babel}
\usepackage{fancyhdr}
\usepackage[mathscr]{eucal}
\usepackage[dvips]{graphicx}
\usepackage{graphics}
\usepackage{color}
\usepackage{epsfig}
\usepackage{amsmath}
\usepackage{amsthm}
\usepackage{amsfonts}
\usepackage{amssymb}
\usepackage{amscd}
\usepackage{bbm}
\usepackage{latexsym}
\usepackage{marvosym} 
\usepackage{pifont}
\usepackage{subfigure}
\usepackage{psfrag}
\usepackage{caption}

\begin{document}

Contact interaction in Quantum Mechanics, Gamma convergence and Bose-Einstein condensation

G.F.Dell'Antonio

Math, Dept  University Sapienza Roma

and 

Mathematics area Sissa

\bigskip

\section{Introduction and summary}

We study a problem in Q.M. that is related to extensions of symmetric operators;  we  make  essential use of Gamma convergence, a variational method introduced by E.de Giorgi in the study of finely fragmented composite media.  

We introduce contact (zero range) interactions in $R^3$, both  "strong" and "weak";  they  are limits, in strong resolvent sense, of hamiltonians with potentials of decreasing range (and different scaling factors in the two cases).  

Strong contact of a particle with two identical particles leads to  the Efimov effect in Low Energy Nuclear Physics. 
The same effect is obtained in the simultaneous weak contact of three (Bose) particles, described by the Gross-Pitayewsii equation, a cubic focusing P.D.E. 

The Gross-Pitayewskii equation is also \emph{the variational equation}  for the energy. 

Contact interactions  describe also the  condensate of \emph{Cooper pairs} in superconductivity. In this case the  equation is still a local cubic focusing P.D.E. but \emph{different from the G-P equation} . 

The Hamiltonian represents local interaction between \emph{two pairs of electrons}  induced by   a singularity of  the Coulomb potential of the nuclei. 

Contact interactions and Gamma convergence have a role also in Solid State Physics  for particles (electrons) which satisfy the Pauli equation and move in  (a narrow neighborhood of ) a lattice  with $Y$-shaped vertices where the  interaction takes place, a model suggested by theoretical analysis and by images taken with  electron microscopes.  The resulting \emph{Fermi sea}  is the filling  of an Efimov sequence of bound states  with $ \frac{1}{\log n}$ rate of decrease of the negative eigenvalues.

Contact interaction may be used also to describe  a \emph{contact (Nelson) Polaron}, both relativistic and non relativistic case (the Nelson Polaron is a particles of mass $m$ in "minimal" interaction with a mass zero field described in second quantization) .

We shall describe elsewhere the last two  structures.   

The mathematical tools we use, a part from Gamma convergence, are rather standard  tools  in the mathematics of Quantum Mechanics.

\section{Contact interactions}

Recall that in Quantum Mechanics the interaction is described by a Schr\"odinger equation  and the hamiltonian is the sum of a kinetic part  $H_0$ (the free hamiltonian, usually a second order partial differential operator,) and a potential part, usually a (negative) function with various regularity properties.

We introduce two different types of "zero range" (contact) interactions, weak and strong,  through the use of of boundary conditions "at contact" (to be defined soon). 

\emph{Formally}  they correspond to placing distributions of different orders as potentials on the boundary $ \Gamma_{i,j}\equiv \{x_i - x_j = 0 \} $  
 
We will prove that the resulting hamiltonians are  limit, in a strong resolvent sense, of sequences of hamiltonians with suitably regular two-body potentials that have support i a neighborhood of $ \Gamma _{i,j}$ of decreasing size.   

The "interaction hamiltonian" for both types of contact must be properly defined; we shall see the role played in this by Gamma convergence, a variational tool introduced six decades ago by E de Giorgi in the contexts of homogenization of finely fragmented materials.   

In both cases a variational method is used to find the minimum of the total energy as a  balance between (positive) kinetic energy and (negative) potential energy.

Contact interactions in $R^3$   are self-adjoint extension  of the symmetric operator represented by the free hamiltonian restricted to function that vanish in a neighborhood    of  the \emph{coincidence manyfold}  $ \Gamma$  

\begin{equation}
 \Gamma \equiv  \cup_{i,j} \Gamma_{i,j} \qquad \Gamma_{i,j} \equiv \{ x_i = x_j \}  ,\;\; i \ne j \quad x_i \in R^3 .
\end{equation}

They are \emph{formally} defined by the boundary conditions  $\phi (X) = \{ \:  \frac {C_{i,j}} { |x_i-x_j|} + D_{i,j}  \quad i \neq j \: \}$

We call \emph{strong contact} the case $C_{i,j} \not=0,\; D_{i,j} = 0  $ and weak contact the case $ C_{i,j} = 0 , \;\; D_{i,j} \not= 0 $.
     
 We  consider only the  case in which  the parameters take the same value for any pair of indices.
 
 The values of the parameters $ C$ and  $D$   characterize the extensions. 
     
 These conditions  were used already in 1935 by H.Bethe and R.Peierls [B,P] (and before them by E.Fermi) in the description of the interaction between proton and neutron. 
They  were later used by Skorniakov and Ter-Martirosian [T,S] in the analysis of three body scattering within the Faddaev formalism. 

 These two types of extension  give  \emph{complementary and independent results} . 
Both results are  \emph{independent and complementary}  to those due to potential of Rollnik class. 

We shall sketch in an Appendix the proof of this complementarity .

 Functions that satisfy strong or weak contact boundary conditions  \emph{are  not in the Hilbert space} $ L^2 (R^3) $ (and therefore not in the domain of the Schr\"odinger hamiltonian $H_0$). 

In the case of strong contact,  integration by parts produces a potential term proportional to $ \delta (x_i- x_j)$. 

Integration by part is \emph{formal} because it requires regularity  of the "dual"  function. The corresponding quadratic form one obtains is not strongly  continuous.  

Also in the case of weak contact a sudden increase of the function at the boundary cannot be obtained with a bona-fide potential.

Therefore the first task is to define   the hamiltonians of strong (and weak)  contact as a self-adjoint operators.  

This can be done for a system that contains \emph{at least three particles}. 

The reason is that we shall  use a method of "virtual fragmentation  and recomposition" that avoids the contribution of single  "delta" singularities (which a placed in a set of capacity zero) while keeping the contribution of the pair   

In the examples we consider three particle systems. 

Later we will prove  that both weak and strong contact hamiltonians are limit, in strong resolvent sense, of Hamiltonians with regular  potentials which  in the case of strong contact 
scale as $ \epsilon \to 0 $ as $ V^\epsilon (|x_1 - x_j) = \frac {1} {\epsilon^3} V ( \frac {|x_i- x_j|}{\epsilon}) $ with$ V(x) \in C_0 \int L^3 (x) $; in the case of weak contact  the scaling is  $ V^\epsilon (|x_1 - x_j) = \frac {1} {\epsilon^2} V ( \frac {|x_i- x_j|}{\epsilon}) $ with$ V(x) \in C_0 \int L^3 (x) $ \emph{and there is a zero energy resonance}. 

In the first case the $L^1$ norm characterizes the extension, in the second case the Rollnik norm

\section{Weak contact}

Consider first weak contact. 

Weak contact  is defined as a self-adjoint extension of $H_0$ (the symmetric operator defined as the free Hamiltonian restricted to functions that vanish in a neighborhood of $ \Gamma$)  for  which  functions in the domain can take a finite value at the boundary. 

The value at the boundary classifies the extensions. 

In $R^3$ weak contact  of two particles requires the presence of a zero energy resonance,  i.e. a solution of the two-body problem which has the asymptotic behavior at large distances $ c \frac {1}{|x| } $ (and therefore it is not in $ L^2 (R^3)) $.  

The relation between the value at $\Gamma_{i,j} $ and the behavior at infinity of the resonance is obvious because there are no other  potentials. .

The presence of a zero energy resonance could  prevent us to  use of compactness in order to prove convergence of the approximating Hamiltonians.

To restore compactness  we require  that also these Hamiltonians have the same zero energy resonance  so that the contribution  of the resonance   can be subtracted away before taking the limit.

In the following we will consider first  the case in which one particle is in weak  separate contact with two particles which are identical bosons (the wave function of the pair is symmetric under interchange).

The symmetric part of the tensor product of the wave function of the two resonances is a  \emph{one-dimensional} subspace $\Xi $ of $L^2 (R^6) $. .

If the potentials were regular this would imply that the hamiltonian has a bound state (and no zero energy  resonances) 

Indeed the inverse of the resolvent  at the origin in momentum space is a two-by-two matrix with zeroes on the  diagonal and has therefore a negative eigenvalue. 

We will show that this is true also for weak contact.  The weak contact hamiltonian for the interaction of a particle separately with two identical bosons has a bound state (and no zero energy resonances).

We will study also a system of three particles which are in mutual weak contact. This system has infinitely many bound states with eigenvalues that scale as $ - c \frac {1}{n}$

We will prove that contact  hamiltonians, strong or weak, are the limit, \emph{in strong resolvent sense} , of hamiltonian with regular potentials (and different  scaling in the two cases). 

Recall that strong resolvent convergence implies convergence  of spectra and of Wave operators (scattering matrices) \emph{but not strong convergence of operators!}. 

\bigskip

\emph{Remark} 

We will also remark, without giving full details, that a system of three particles  which mutually interact through a weak contact hamiltonian has as semiclassical  limit 
\emph{the classical  three-body newtonian system} (and the bound states correspond to the periodic orbits).  
 \bigskip
 
 ................................... 

\bigskip

\emph{Point interaction} [A] describes a particle of mass $m$ in weak contact with a point; as in any weak contact there is a zero energy resonance. 

The study of the resolvent in the case of point interaction is difficult [A] because the singularity of the free resolvent at zero momentum "interferes" with the singularity due to the zero energy resonance. 

This case can be studied [G,Y] as limit of two body interaction when the range of the interaction potential goes to infinity (a resonance) and the support of the interaction goes to zero.

It can also be studied  as  the system of weak separate contact of two identical Bose particle with a third particle \emph{if seen at a very large scale}.

At this scale only the kinetic energy of the barycenter is retained and the product of the two bound states  has in the center of mass coordinates the properties of a zero energy resonance

At this scale the system represents therefore a point interaction.

This method of looking at a distance (so that details are not seen) is a "poor man version" of the variational method we shall use, Gamma convergence

As in [A],  in "separating "  the singularity at zero momentum coming the kinetic term from that due to the zero energy resonance. 

\bigskip

............................

\bigskip
 
\section{Strong contact} 

Consider next strong contact.  

Strong contact is not defined for a two-particle system.

For a three particle system strong \emph{separate} contact is defined as the self-adjoint extension  that has in its domain functions that have a $ \frac{c}{ |x - x_j|}$ behavior at the boundary, where $ j = 1,  2  $ (notice that functions with this behavior at the boundary can be in the Hilbert space $L^2 (R^6)$. < 

Again different extensions are classified by the value of the constant $c$. 
 
We give soon the  mathematical construction. 

 Making use of Gamma convergence we prove that  a system of three identical particles one of which interacts separately by strong contact with  other two   is described, if the interaction is sufficiently strong, by a self-adjoint Hamiltonian that has infinitely many bound states with energies that scale approximately as $ - \frac {c}{ \sqrt {n}} ,\;\; c > 0 $ (Efimov effect). .
 
We will show that the hamiltonian is the limit, \emph{in strong resolvent sense}, of hamiltonians with potentials that have supports which shrink to a point  (we will more precise later) 

We will later prove that the same Efimov effect occurs also  in the case of simultaneous  weak contact of three identical bosons .

\bigskip

\emph{Remark 1}

We will  show that i similar chain of arguments gives the bound states of two pairs of electrons ( \emph{ Cooper pairs}) which are both in weak contact with  " a phonon" .

They  are described by a \emph{different} cubic focusing P.D.E.

\bigskip

......................
 
\emph{Remark 2}

We stress again the method we use is successful  for a system of \emph{two separate strong contacts}. 

It  is a variational method that use fragmentation and homogenization; is is not a \emph{separate regularization of the two "delta" potentials}. 

\bigskip

..........................

\bigskip

We are  looking for extensions of $ \hat H_0 $, the symmetric (but not self-adjoint) operator defined as the free Hamiltonian of a three-body system restricted to functions that vanish in a neighborhood of the contact manyfold $\Gamma $.

In Theoretical and Mathematical Physics the  interest in the subject was renewed by recent advances in  Low  Energy Physics and by the flourishing of research on ultra-cold atoms interacting through potentials of very short range.  

Strong contact of two particles is not defined (it would lead to divergencies).

In what follows we analyze first the case of three particles (bosons) one of which is in separate strong contact with the other (identical) two.

Afterwards  we will describe the case of separate weak contact of one particle with other two and of joint weak contact among three identical particles.

If the interaction is sufficiently strong both  strong contact and joint weak contact produce   the \emph{Efimov effect} [E] i.e. the presence of an infinite number of bound states with eigenvalues that converge to zero as $ \frac {-C}{\sqrt n} $. 

All particles are bosons and for simplicity all particles are identical. 

The quadratic form of the free hamiltonian is a  positive form with domain the space of absolutely continuous functions.

The delta distribution defines a bounded negative bilinear form in this space. 

Therefore \emph{if the sum defines  a self-adjoint operator}, this operator is bounded below.  

But the quadratic form defined by the delta function is not strongly closed; it remains to be proven that this self-adjoint operator exists. 

We shall prove that in presence of a third particle  this operator exists and is self-adjoint; we shall determine its spectrum. Gamma convergence will be needed for this purpose. 

Later we will do the same for weak contact of a particle with two identical particles  and for the simultaneous weak contact of three identical particles. 

 \bigskip

We must find an operator that is  "almost" as delta function. 

We are not able to do this, unless making use of non-standard analysis 

We use then a different strategy:  we take advantage of the regularity of the wave function of the non interacting particle and  integrate  by parts with respect to its  coordinates.

The invariant   formulation of this operation is realized through the  \emph{ the Krein map } $ { \cal K}$ ,  to a space of more singular functions.

We  call thus space  "Minlos space" $ {\cal M}$ because the idea  came from reading [ M1].

The map "regularizes more"  the potential  term  than the kinetic energy. 

The map is "fractioning"and "mixing" in a suitable sense. 

The emphasize that the Krein map \emph{is only a tool}; it serves the purpose of clarifying the stricture of the boundary condition that we  \emph{formally} attributed to a delta function potential. 

The space  $ {\cal M} $ is obtained from  $L^2 (R^9) $ acting with $(H_0 ) ^{-\frac {1}{2}}$; since we are going to invert the map, we can assume that  it acts as identity on the complement of range of $H_0$). 

We can alternatively choose the operator $ H_0 + \lambda, \;\; \lambda > 0 $ and let $ \lambda \to 0$ at the end.
 
For the sake of simplicity of notation we take $ \lambda = 0 $. 

Since it serves to clarify the condition at the boundary,  \emph{the Krein map acts DIFFERENTLY  on the kinetic terms and on the potential term i.e. on operators and on weakly closed quadratic forms}.   

On the potential part (which is only defined as quadratic form) it acts as $ \delta(x_i - x_j)   \to  (H_0  )^{-\frac {1}{2}} \delta (x_i -x_j)  (H_0  )  ^{-\frac{1}{2} } $. 

 On the free Hamiltonian (an operator) the Krein map  acts  as $ H_0  \to (H_0 )   ^{\frac {1}{2}}$ .
 
 Regarded as bilinear forms $H_0$ and the delta commute;  they both are singular elements of the abelian algebra $ { \cal A }$ of functions of momenta. 
 
Therefore the map  can be written  formally  as $\delta  \to  \delta  (H_0 )^{-1}  $. 

Our approach  is therefore in the path followed  by Birman, Krein and Visik [B][K] for the study of self-adjoint extensions of positive operators (this is the reason for the choice of the name).  The main difference is that we work with quadratic forms [A,S] [K,S] 

On the kinetic energy the map can be written as $ H_0 \to (H_0  )^{-\frac {1}{4}} H_0  (H_0  )^{-\frac {1}{4}}.$

Therefore both the free hamiltonian and the "generators" of the maps belong to the  algebra $ { \cal A}$. 

It is easy to verify [M1] that in $ {\cal M}$   the "potential" has in position space in both channels the form $ -\frac {C}{|x| }+ B  $  where $ B$ is a \emph{positive} bounded operator with kernel that vanishes on the diagonal and the constant $C$ depends on the coupling constant and on the mass of the particles. 

Results of this type  in $ { \cal M}$  are also obtained in   [M2] and in [C] in another context.
 
Therefore in  $ {\cal M} $ the potential term is a bona-fide operator.

 In $ {\cal M}$ the kinetic energy  is a pseudo-differential operator of order one. 

The singularities at the origin (in position space) of the kinetic and the potential terms are \emph{homogeneous} .

Therefore [D,R][l,O,R] there are values $C_1$ and $C_2$ of the positive constant   $C$ ( that depend on the masses and on the coupling constant) such that for $ C < C_1 $ their sum is a positive weakly closed quadratic form (and represents therefore [K] a positive self-adjoint  operator)  

For $C_1 \leq C < C_2 $ there is in ${ \cal M}$ a continuous family of self-adjoint operators $H_{C, \alpha} $ each with one negative eigenvalue $ \lambda  (C, \alpha) $.

For $ C \geq C_2 $ there is in ${ \cal M}$  a continuous family (parametrized by $ \alpha $) of self--adjoint operators unbounded below each with a sequence $ \lambda_n (C, \alpha) $ of negative eigenvalues with the asymptotics $ \lambda_n (C, \alpha)  \simeq  - b(C. \alpha) log n $ 

 These results are  derived [D,R][l,O,R] using the Mellin transform and properties of the Bessel functions. 

The  Mellin transform "diagonalizes" the sum of the kinetic and  potential terms.  

This can be proved by explicit computations  [C] [M1] but can be seen also as a consequence of the fact that both forms belong to the abelian algebra $ {\cal A}$ and the Mellin transformation   corresponds to a change of coordinates. 

 We must now  come back to the original physical space. This is done by using Gamma convergence. 
 
 Notice that we do not make a point-wise inversion of the Krein map. 
 
 We go back to physical space by a variational argument (as in homogeneization) neglecting "detains" (e.g.  the \emph{separate}  singularities of the two delta functions) .

The Krein map $ {\cal K} $ is \emph{fractioning} (the target space is a space of less regular functions) and \emph{mixing} (the square root is not diagonal in the channels).

This suggests the use of Gamma convergence [Dal] , a variational method introduced sixty years ago by E. de Giorgi and originally used for the study of "homogenization" of finely structured materials.

Gamma convergence selects the infimum of an ordered  sequence of strictly convex quadratic forms bounded below that are in a compact domain of a topological space   $ Y$ .

Since the interaction potential is rotational invariant, only the  s wave is affected so that  all  forms are \emph{strictly convex}. 

The Gamma limit $ F(y)  $ is the quadratic form characterized by the following relations

\begin{equation} 
\forall y \in  Y, y_n \to  y, F(y) = lim inf F(y_n) \;,\; \forall x \in  Y _n \forall  \{ x_n  \} : F(x) \leq  limsup_{n} F_n(x_n) 
\end{equation},

 The first condition implies that $F$ is a common lower bound for the  $F_n$ , the second implies that the bound is optimal. 

The condition for existence of the Gamma limit is that the sequence be contained in a compact set for the topology of $Y$ (so that a Palais-Smale converging sequence exists). 

In our case the topology is the  Frechet topology defined by the Sobolev semi-norms and compactness is assured by the absence of zero energy resonances. 

Therefore in our case the Gamma limit exists. 

By a theorem of Kato [K] the "lowest form" admits strong closure and this defines a self-adjoint hamiltonian.

\emph{This is the hamiltonian that represents separate strong contact of a particle with two identical particles}. 

 If  $ C_1 \leq  C < C_2 $ this Hamitonian has a bound state, if  $ C\geq C_2 $ it has an Efimov  sequence of bound states with eigenvalues that scale as   $-c \frac {1}{ \sqrt {n}}$..

The  eigenvectors can also be given explicitly and therefore the model is in this sense \emph{completely solvable}. 

It is possible to prove [D,R] that the moduli of the eigenfunctions have the form  $ \frac {1}{ |x_i -x_j| log (n |x_i-x_j|)} $. 

The orthogonality is due the fact that the phases increase linearly with  $n$. 

Since the eigenfunctions of the "excited states"  are complex valued, only the ground state can be obtained by Morse Theory. 

For the other one must use Index Theory or the symplectic version of Morse Theory [Ek].

\bigskip

\emph{Remark 1 }

One may wonder what happens to the quadratic forms that are not an extremal and therefore are not strongly closable. 

They correspond to different boundary conditions at contact; this alters the order of the quadratic forms and the Gamma limit is different. 

We have used the free Schr\"odinger Hamiltonian to construct the Krein map; we could instead use the magnetic Schr\"odinger Hamiltonian and we would reach other extensions. 

\bigskip

\emph{Remark 2} 

The procedure we have followed is such that at the end of the process the kinetic energy is not changed while the potential term is \emph{regularized}; it is  therefore a sort of \emph{regularization  of the interaction} but notice that the method is variational and non perturbative and the "regularization" is done for the pair of potential and cannot be done for a single "delta" potential.  .  

,

\bigskip 

.................................

\bigskip

\section{Strong resolvent convergence of approximating Hamiltonians}  

We shall now prove that the strong contact hamiltonians are limit \emph{in strong resolvent sense} of Hamiltonians of three body systems in which one particle interacts separately  with the other two through potentials of class $L^1(R^3) $  which scale as $ V^\epsilon(x_i - x_j) =   \frac {C}{\epsilon^3}V( \frac {|x_i -x_j|}{ \epsilon} )   $.

The $ L^1 $ norm of the approximation hamiltonian is invariant under scaling and  classifies the extensions: it gives the coefficient of the  $ delta $ function that \emph{formally}describes the strong contact potential. 

Such Hamiltonians do not have zero energy resonances; therefore the quadratic forms  belong to a compact set in the topology given by  Sobolev semi-norms. 

The sequence is strictly decreasing as a function of $ \epsilon$ and it  is bounded below by the quadratic form of the strong contact interaction.  
 
Every strictly decreasing sequence in a compact space has a unique limit if it is uniformly bounded below.  

The limit is contained in the image under the inverse of the Krein map of the limit set;  the sequence has  therefore the same Gamma limit as the quadratic forms that we have described previously. 

Therefore the $ \epsilon$-hamiltonians Gamma converge  to the hamiltonian of strong contact interaction. 

 Gamma convergence implies strong resolvent convergence; therefore the  hamiltonians $ H_0 +  V^\epsilon (x_0 - x_1 ) +V^\epsilon (x_0 - x_2 ) $ converge \emph{in the strong resolvent sense} to the Hamiltonian of separate strong contact of a particle with two identical particles. 

 Notice that strong resolvent convergence implies convergence of spectra and of Wave Operators but \emph{does not imply convergence of quadratic forms}: it only implies convergence of those sequences  that remain uniformly bounded [Dal].

 Remark also that the method we follow is variational and therefore \emph{no estimates} can be given of the convergence as a function of the parameter $\epsilon $ . 
The result \emph{cannot be obtained in perturbation theory}. 

If $C$ is large enough this leads to an Efimov sequence of bound states. 

Gamma convergence is stable under the addition of regular potentials; the proof makes use of a formalism introduced by T. Kato and improved by Konno e Kuroda [K,K] . We will sketch it briefly in an Appendix. 

\section{Efimov effect in Low Energy Physics} 

We have seen that in order to have a self-adjoint operator describing the strong contact of two particles, a third non interacting particle is needed. 

This is the case for a system of two species of particles (e,g, protons and neutrons) in which members of each specie don't interact  among themselves and interact  with members of the other specie. 

Therefore  a three-body problem is sufficient to describe the effect.

In Low Energy Physics the interactions between particles  can be described approximately by Hamiltonians with two body potentials of very short range.

We have seen that for a three body system  these hamiltonians have as limits in a strong resolvent sense, when the range of the potential goes to zero, a strong contact hamiltonian. 

In this limit the system has an Efimov sequence of bound states and the sequence of eigenvalues has  rate $- c \frac {1}{\sqrt n} $  where the constant $c$ depends on the nuclei. 

This Efimov property has been verified experimentally in the sense that it is approximately true  in several composite nuclei for a first few eigenvalues.

Since in reality the interactions within the nuclei are of finite (however small) range, and therefore strong contact of three particles is an approximation, one should not expect to see the full Efimov series and the  presence of a few elements is significant.  

\bigskip

\section{Weak contact of  a particle separately with two identical particles}

We again treat separately the two contacts;  taking advantage from the presence of the other particle we introduce  a Krein map $ {\cal K}$  and the Minlos space $ { \cal M}$ defined as before. 

We use in this case a different Krein map; we now act in the same way on the kinetic and the potential parts; also in this case  the map is induced by $ H_0 ^{-\frac {1}{2}}$.

The Krein map is again mixing and fractioning.  

This time the  potential is   $ - C  log |(x -x _1) $ and the kinetic term is still  $ \sqrt {H_0} $

In $ { \cal M}$  there is a family of  quadratic forms  symmetric and bounded below. 

Return now to "physical space". 

In this space there is a family of quadratic forms weakly closed. 

The interaction is still only in the angular momentum zero channel and in this channel the quadratic forms are   strictly convex. 

In the three-particle system there are still no zero energy resonances.

Therefore we can use Gamma convergence; the "minimal form" can be closed strongly and the result is a self-adjoint operator, the Hamiltonian of our system 

This self-adjoint operator  has a bound state  with  eigenvalue  proportional to the value at contact of the functions in the domain of the chosen extension.   

Notice once again that we have not given meaning to the \emph{separate} two-body interactions.

\section{Joint weak contact of three identical particles: the Gross-Pitayewskii equation}.

To derive the G-P equation we consider we consider   mutual weak contact of a system of three identical "particles" in $R^3$. 

We take the particles to be \emph{bosons} (i.e, the wave function is symmetric under interchange of the coordinates). 

We make again use of the Krein map to define the operator in "physical space".  

An additional strong constraining potential is needed to keep the system confined. 

Notice (see Appendix) that weak contact and regular  (Rollnik type) potentials  lead to \emph{indepedent and complementary} effect [K,K].

\bigskip

\emph{Remark} 

Also for weak contact it is possible to show that the resulting hamiltonian is the limit, when $ \epsilon \to 0$ of the  hamiltonian with the approximated potentials. 

\bigskip

We prove now that weak simultaneous contact of three identical particles leads to the Gross-Pitayewskii equation. 

Recall that, in the case of strong separate contact, in the space  $ {\cal M}$ the potential term contains  a \emph{positive} term that we have called $B$.  

We prove  that in the case of mutual weak contact of three identical particles \emph{this term is not present}.

Therefore now the kinetic energy term and the potential energy terms have the same behavior under scaling (not surprisingly since the  Hamiltonian of weak contact is covariant under scaling).

We will   use the Birman-Schwinger formula for the difference of the resolvent of the  hamiltonian for  \emph{simultaneous weak contact}  of three particles and the resolvent of the strong separate case. 

The B-S formula for regular potentials is 

\begin {equation} 
\frac {1}{H_2  - z} -  \frac {1}{H_1 - z}  =  \frac {1}{H_2  - z} K_{1,2} \frac {1}{H_1  - z},  \quad  K = U^{\frac {1}{2} } \frac {1}{H_1 - z}  U^{\frac {1}{2} } \quad U = -(H_2- H_1) 
\end{equation} 

where $ H_2 , \;\; H_1  $ are self-adjoint operators, and $ z$ is chosen out of the spectrum of both operators and $ K$ is called Birman-Schwinger kernel.
 
This formulas are valid for the approximating hamiltonians. 

We  take for $H _2$ the Hamiltonian of joint  weak contact and for $H_1 $ the hamiltonian of strong  separate contact. 

We already proved for $H_1$ strong resolvent convergence for an $\epsilon-$ sequence of approximate hamiltonians as $ \epsilon \to 0$ . 

Take $ \epsilon $ finite and consider the contribution  to $ \frac {1}{ H_\epsilon -z} $ coming from  those terms in the perturbative in which all three potentials contribute. 
 
In these contributions  we can "take away" the factor $ \epsilon^{-2}$ from one of the potentials  and attribute a factor $ \epsilon ^{-1} $ to each of the other two. 

Together with the kinetic energy the limit $ \epsilon \to 0$ provides in physical space the resolvent of $H_1$ .

The remaining term comes from the interaction of two out of three of the particles. 

For the separate interaction of one particle with two other ones in physical space quadratic form convergence and resolvent convergence imply each other.  

It easy to see that in $ \cal M$ in the limit $ \epsilon \to 0$ their contribution cancels, in the quadratic form,  the term  $B$ that was present  in the case of separate strong contact. 

Indeed the contribution  $B$ was due to the separate weak contact of one particle with a pair of identical particles.

We can now take  the limit $ \epsilon \to 0$ and invert the Krein map.. 

To prove that $H_2$ is a self-adjoint operator, we use once again Gamma convergence

Since $B$ is positive, the  minimum of the energy functional is provided   by the simultaneous weak contact of three particles \emph{AND NOT BY STRONG CONTACT OF TWO
 Of THE PARTiCLES}. 

We proved that $ H_1$ has an Efimov sequence of bound states that scale as $ - c \frac {1}{\sqrt {n}}$ where  the constant $c$ depends of the extension. 

Each of the infinitely many bound states in the case of joint weak contact is lower that the corresponding state of separate strong contact. 

The lowest energy state of the system is the ground state for weak simultaneous contact interactions.

As before, this self-adjoint operator is the limit, in strong resolvent sense,  of the approximate hamiltonians with  a zero energy resonance

In the case of simultaneous weak contact of three particles the hamiltonian is the limit in strong resolvent sense of the sum of the free Hamiltonian and the product of three  attractive two-body potential that scale as $ V^\epsilon ( x_i-x_j) = \frac{1}{\epsilon^2} V ( \frac {x_i -x_j}{\epsilon}) $ \emph{and have a zero energy resonance}.

The system is a three-body system and the interaction is of simultaneous contact. We use translation invariance and choose the origin of the  coordinates in the point where the interaction take place. 

The ground state $\phi_0 (x)$  s unique and therefore its wave function can be chosen positive. 

The interaction is "concentrated  in a point" and therefore the potential due to the interaction with the two other particles is $- g \phi_0(x)  ^2$.

Each particle satisfies therefore  the cubic focusing Gross-Pitayewskii  equation 

\begin{equation} 
i \frac {\partial }{\partial t} \phi _0 (x) = -\Delta \phi _0 (x) - g \phi _0 (x)^3 \;\;\;\; x \in R^3 
\end{equation} 

 where $g$ depends on the extension.

Taking the scalar product with a wave function  the energy functional is 

\begin{equation} 
E(\phi) = ( \phi , H_0 \phi) - g \int  \phi (x) ^4 dx
\end{equation}

We shall call it Gross-Pitayewskii energy functional; the G-P equation is the variational equation  for the G-P energy functional.

We already remarked that in order to keep the condensate confined a strong confining potential is needed , and that weak contact interactions and regular Rollnik class potentials have \emph{ independent and complementary} effects. 

\bigskip

We have proved

\bigskip

\emph{Theorem }

The Bose-Einstein condensate  is  a gas of separate triples of particles in mutual weak contact  that are held together by strong constraining forces.   

Here "particle" is synonimous of wave function.
 
If the constraining forces act separately on the triples the   ground state  is a tensor product state. It is the tensor of the ground states  of the Gross-Pitayewskii energy functional in which the potential of the forces is added.. 

\bigskip 

....................

\bigskip

\emph{Remark} 

If one regards as "building blocks" the density  matrices ,  instead of considering the product of three wave functions one looks at the product of two density  matrices (each density matrix is a trace class bilinear form $ \rho(x,y) $ ).

The interaction is now a point interaction of two density matrices.   . 

It is therefore represented  by  a quadratic map $ R(x) \to g R (x) ^2 $ where $R (x) = \rho(x,x) $.

If a description of the system with density matrices  $ \rho (x)$ is introduced,   the potential part of the energy functional is $ - g \int \rho (x)^2 dx $.

This leads to the interpretation of the Bose-Einstein condensate  as a \emph{"gas of pairs of identical density matrices" } in strong contact interaction; the gas is  held together by  strong constraining potentials. . 

The correlations are present for the system with the approximating potentials $V^\epsilon $ but one must take into account that the method  we use is variational and the convergence we proved need not be modulo terms that vanish as some power

 of  $ \epsilon $

\bigskip

..................................

\bigskip

\section{Condensation and superconductivity} 
 
The regime of condensation that  we have described so far may be called \emph{three-partice regime} because three particles are  bound by  mutual weak contacts.
 
The three-particle system is stable and the ground state satisfies the Gross-Pitayewskii equation.

There is a different regime of condensation.

In this different regime there are still zero energy resonances but now \emph{there is no joint three-particle weak contact} . 

This condensation is  related to superconductivity.  
 
The setting is commonly described in Solid State Physics as a condensate of \emph{Cooper pairs}; the mechanism for their formation has been described by Bardeen-Cooper-Schrieffer [B,C,S].  

In a crystal the interaction of the nuclei with the electrons causes a slight modification in the position of the nuclei of the crystal.

Cooper showed that in  a metal this deformation of the lattice of ions in a crystal due to  the attractive  interaction may give rise  a  bound state of two electrons with different spin orientation. 

This bound state  is usually called   \emph{phonon};  phonons have the properties of bosons. 

At the atomic scale this bound state can be very extended . 

Two pairs  of electrons may "interact"  by the fact that they "share a phonon" (the same deformation of the lattice of ions is felt by two  pairs of electrons with different spin orientations) .

This can be formalized  as a weak \emph{ repulsive} contact of two pairs of electrons which are in a bound state produced by the deformation of the lattice of the ions; one may consider this an interaction of comparatively  large and \emph{negative}  scattering length. 

This process is described as  the  formation of Cooper pairs;  Cooper pairs have the statistics of boson.

For its \emph{formal analogy}  with the Bose-Einstein condensate, the formation of this  condensate is often referred to as  B.C.S to B.E. transition. 

Notice that the formation of the pairs is attractive (  positive large scattering length) whereas the interaction between pairs is repulsive (negative large scattering length). 
.
We summarize this analysis in 

\bigskip

\emph{Proposition }

The  \emph{condensate of Cooper pairs} is a condensate \emph{pairs of electron }.

It is produced by a weak  \emph{repulsive}   contact of two pairs of particles (electrons) which are in a bound state.  

The bound state  can be considered as the result of weak attraction of two pairs. 

The repulsion between the pairs is due to "the exchange of a {quasi particle}" (a "phonon").

\hfill $ \diamondsuit $ 

\bigskip

The resulting energy functional is not the G-P functional  but rather the functional 

\begin{equation} 
( \phi, H \phi ) = \int ( \phi (x), H_0  \phi  (x) dx - c \int  ì\phi (x)| ^2  \delta (|x_1 - x | ) |\phi (x_1)| ^2 | dx dx_1
\end{equation}

We shall call this the "weak" Gross-Pitayewskii energy functional or  BCS functional. 

The localization is weaker (the delta function in on the modulus of the distance))  

Notice that this functional puts in evidence the \emph{interaction  between pairs}

 The functional   is \emph{local}  ( the  $\delta |x- x_1| $ localizes  the interaction)  but does represent  contact. 

The corresponding P.D.E. is the "weak" Gross-Pitayewskii equation (or B.C.S equation) 
 
 \begin{equation} 
  \frac {\delta}{ \delta t } \psi (t,x)  =  H_0 \psi (t,x) - 2 c \int \psi (t,x) \delta (|x-x')| | \psi(t,x')|^2  dx' 
 \end{equation}  

still a local cubic focusing P.D.E. but with a weaker interaction (the delta function is only in the modulus of the distance)

\bigskip

....................

 \section{Semiclassical limit}  

One verifies easily  that weak contact provides Coulomb interaction between the barycenters of coherent states. 

In this sense the Newtonian three body system can be regarded as semiclassical limit of the quantum three body problem with a simultaneous weak contact. 

The Efimov states of weak contact of three particles are the counterpart of the closed orbits of the classical three-body problem. 

Both systems have an infinite number of periodic solution (resp. bound states) ad the discrete spectra correspond (eigenvalue of the bound states on one side,  energy  of periodic orbits on the other side) 

Therefore this "weak form"  of the semiclassical limit is satisfied.

 One has also a "W.K.B."   correspondence.
 
 For a WKB analysis of the cubic defocusing cubic equation see [Ca] [ G]

For the focusing cubic equation (related to simultaneous weak contact of three bodies)  there are no zero energy resonances and  the analysis in [Ca] holds (the WKB semiclassical limit can be taken for initial data in  ${\cal H}^1$). 

For the discrete part of the spectrum the weak contact among three particles is in some sense  "completely solvable" : it allows a complete description of the eigenvalues and of the (generalized) eigenfunctions.

A more direct  W,K,B approach to the semi-classical limit is the following.

Without changing the dynamics we can use the Hamiltonian $ H_0 + \lambda , \;\; \lambda \in R $

For $ \lambda \to \infty $ the Krein map can now be related to the semiclassical limit. The space $ { \cal M}_{\lambda} $ for $ \lambda \simeq  \hbar^{-1}  $  can be regarded as semiclassical space. Setting $ \frac {1}{\sqrt{\lambda}}= \hbar  $, a part from an irrelevant constant one has in $ { \cal M}_\lambda $  to first order in $ \frac {1}{\lambda} $ the hamiltonian of the three-body newtonian system.

Notice that in the semiclassical limit the free hamiltonian is scaled by a factor  to $ \hbar ^{-2} $ and the Coulomb potential is scaled by a factor $ \hbar^{- 1} $ 
If we identify the radius of the potential (the parameter $ \epsilon$ ) with $ \hbar $ (both have the dimension of a length)  the limit $ \hbar = \epsilon \to 0$ gives weak contact interaction at a quantum scale, Coulomb interaction at a semiclassical scale. 

\bigskip

\emph{Remark }  

In thus respect we mention that in a recent preprint [arXiv:2002.11638 v1 ) T.Kappeler and O.Topalov [K,T] have proved a Arnold-Liouvillle type of Theorem for the Focusing  NLS equation for periodic boundary conditions. 

With some care contact interactions can be defined also for the periodic case (the constant solution takes the place of the zero energy resonance). 

The result of Kappeler and Topalov can be regarded as the classical counterpart of the results described here. 

\bigskip

.........................

\bigskip

 \section{APPENDIX: Strong contact, weak contact and regular interactions have complementary effects}

\bigskip

\emph{Theorem A}

\bigskip

In three dimensions for   contact interactions and  weak-contact interactions  contribute \emph{separately} and \emph{independently}  to the spectral properties and to the boundary conditions at the contact manyfold.

Contact interaction contribute to the Efimov part of the spectrum and to the T-M boundary condition
$ \frac {c_{i,j} }{|x_j - x_i|} $ at the boundary $ \Gamma \equiv \cup_{i,j} \Gamma_{i,j} $. 
Weak-contact interactions contribute to the constant terms  at the boundary and may contribute to the  (finite) negative part spectrum.

\hfill $ \diamondsuit $

\bigskip

For an unified presentation (which includes also the proof that the addition of a regular  potential does not change the picture) it is convenient to use a symmetric presentation due to Kato and Konno-Kuroda [KK]  (who generalize previous work by Krein and Birman)  for hamiltonians that can be written in the form

 \begin{equation}
H= H_0 + H_{int} \qquad H_{int} = B^* A   
\end{equation}

 where $B, \;A$ are densely defined closed operators with $ D(A) \cap D(B) \subset D(H_0) $ and such that, for every $z$ in the resolvent set of $H_0$, the operator $A \frac {1}{ H_0 + z} B^* $ has a bounded extension, denoted by $Q(z).$
 
We give details in the case $N=3$.

Since we consider the case of attractive forces, and therefore negative potentials, it is convenient to denote by $ - V_k( |y|) $ the two body potentials. The particle's  coordinates are $x_k \in R^3 , k=1,2,3$. We take the interaction potential to be of class $C^1$  and set 

\begin{equation} 
V^\epsilon (X) = \sum_{i \not= j} [ V_1 ^\epsilon (|x_i -x_j|) +V_2 ^\epsilon (|x_i-x_j|)  + V_3 ^\epsilon (|x_i-x_j|)]
\end{equation} 

where $ V_1$ and $V_2$ are negative and $V_3(|y) $ is a regular potential.

For each pair of indices $ i,j$ we define   $ V_1^\epsilon (|y|) ) = \frac {1}{\epsilon^3}  V_1(\frac{|y|}{\epsilon})$ and $ V_2 ^\epsilon (|y|) = \frac {1}{\epsilon^2}  V_2 (\frac{ |y|}{\epsilon})$.We leave $ V_3 $ unscaled. 

 We define $ B^\epsilon = A^\epsilon = \sqrt { -V^\epsilon}$. For $ \epsilon > 0$ using Krein resolvent formula one can give explicitly the operator $B^\epsilon $ as \emph{convergent power series} of products of the free resolvent $ R_0 (z) , Re z > 0 $ and the square roots  of the sum of  potentials $ V_k  ^\epsilon \ \   k=1,2,3$.
 
 One has then for the resolvent $ R(z) \equiv \frac {1}{ H +z } $  the following form [K,K] 

\begin{equation}
R(z) - R_0 (z) = [R_0(z) B^\epsilon ] [1 - Q^\epsilon(z )]^{-1} [ B^\epsilon  R_0 (z)]   \qquad z > 0
\end{equation}  

with

\begin{equation}
R_0 (z) = \frac {1}{ H_0 +z} \qquad Q^\epsilon (z) = B^\epsilon \frac {1}{ H_0 + z }   B^\epsilon 
\end {equation}

\emph{Proof of Theorem A} 

We approximate the zero range hamiltonian with the one parameter family of hamiltonians 

\begin{equation} 
H_\epsilon =  
 H_0 + + \sum_{m,n}V^\epsilon (|x_n -x_m|)  \;\; n \not=  m , x_m \in R^3 
 \end{equation}

The potential is  the sum of three terms 

\begin{equation} 
V^\epsilon (|y|) = \sum_{i=1}^3  V_i ^\epsilon, \quad   V_1^\epsilon (|y|) =  \frac {1}{\epsilon^3}  V_1(\frac{|y|}{\epsilon}), \; \;  V_2 ^\epsilon (|y|)=  \frac {1}{\epsilon^2}  V_2 (\frac{ |y|}{\epsilon} )
\end{equation} 

(we omit the index $ m,n $) . All potentials are of class $ C^1$. The potential  $ V_3 $ is unscaled. 

\ 
Define 

\begin{equation}
 U^\epsilon (|y|) = V^\epsilon_2 +  V_3 
 \end{equation}
 
  If $ \epsilon > 0$  the Born series converges and the resolvent can be cast in the Konno-Kuroda  form,  [K,K] where the operator $B$ is given as (convergent) power series of convolutions of the potential $ U^\epsilon $ and $V^\epsilon_1 $  with the resolvent of $H_0$. In general 

\begin{equation} 
\sqrt {V_1^\epsilon  (|y|) + U ^\epsilon (|y|)} \not= \sqrt {V_1^\epsilon  (|y|) }+ \sqrt {U^\epsilon (|y|)  } 
\end{equation} 

and in the Konno-Kuroda  formula for   the resolvent of the operator $H_\epsilon$ one loses separation between the two potentials $V^\epsilon _1 $ and $ U^\epsilon  $.

Notice however that, if $V_1^\epsilon $ and $U^\epsilon$ are of class  $ C^1$ ,   the $L^1 $ norm of $U ^{\epsilon}$ vanishes as $ \epsilon \to 0$ uniformly on the support of $ V^\epsilon _1$.  By the Cauchy inequality one has 

\begin{equation}
lim_{\epsilon \to 0 }    \| \sqrt { V_1^\epsilon  (y)} .\sqrt { U^\epsilon  (y) }\|^1    = 0 
\end{equation} 

Therefore if the limit exists  the strong and weak contact interactions act independently.

\hfill $\heartsuit $

In the same way one proves that the addition of a "regular"potential (e.g. a potential of Rollnik class) leads to independent and complementary results. 
 
Let the regular part be a smooth two body $V(x) \in L^1\cap C $.

Following the same steps that led to the proof of Theorem 1 one proves  that this  interaction \emph{contributes separately} to the spectral structure of the hamiltonian.

\bigskip  
 
\hfill $ \heartsuit $

\section{References}

[A]  S.Albeverio,R. Hoegh-Krohn, Point Interaction as limits of short range interactions on Quantum Mechanics   J. Operator Theory   6 (1981) 313-339 

[A,S] K.Alonso, B.Simon  The Birman-Krein-Visik theory of self-adjoint extensions.  J.Operator Theor"y 4 (1980) 251-270 b  

[B] M.Birman On the theory of self-adjoint extension s of positive definite operators Math. Sbornik. N.S. 38 (1956) 431-480 i  

[B,M,N] J.Brasche, M.Malamud, M.Neihardt , Weyl functions and spectral properties of self- adjoint extensions Int Eq. Op. Theory 93 (2002) 264-289

[B,O,S] N.Benedikter, G. de Olivera, B.Schlein  Derivation of the Gross-Pitaiewsii equation Comm Pure Appl. Math 68 ( 2015) 1399-1482 

[B,P] H.Bethe-R.Peierls The scattering of Neutrons by Protons. Proc of the Royal Soc. of London Series A 149 (1935) 176-183

[C] M.Correggi, G.F.Dell'Antonio, D.Finco, A.Michelangeli, A.Teta. A class of hamiltonians for three fermions at unitarity Journal Math. Phys. Anal. Geom   18 (2015) 32-60

[Ca]  R.Carles On the semiclassical limit forth enon-linear Schr\"odinger equation arXiv:math/o612518  v 1 

[Dal]  G.Dal Maso Introduction to Gamma-convergence Progr Non Lin. Diff Eq. 8, Birkhauser (1993)  

[D,R] S.Derezinky, S.Richard  On the Schr\"odinger operator with inverse square potential on the positive real line. Arkiv  1604 03340

[E] V.Efimov Energy levels of three resonant interaction particles. Nucler Physics  A 210 (1971) t 57-186

[Ek] I.Ekeland  Une thèorie de Morse pour le systemes hamiltoniens C.R. Acad Sc Paris  I 298  (1983) 117-120 

[E,S,Y] L.Erdos, B. Schlein, H T Yau  Derivation of the Gross-Pitaevsii equation for the dynamics of a Bose-Einstein condensate Ann. of Math. 172 (2010) 291-370

[G] E. Grenier  Semiclassicl limit of th enon-linear Schr\"odinger eqiation in small time  Proc.Am. Math. Soc. 126 (1998) 523-530m 

[G,Y]  A. Galtbayar, K.Yajima   On the approximation by regular potentials of point interactiion ArXiv:1908.02936   2//8/2019n

[K] M.G.Krein  The theory of self-adjoint extensions of semi-bounded hermitian transformations Math. Sbornik   N.S. 20 (1962)1947 431-495

[K,K]  R.Konno, S.T. Kuroda   Convergence of of operators J, Fac, Sci.  Univ. Tokio  13 (1966) 55-63v

[K,T].Kappeler and O.Topalov   Arnold-Liouville  theorem for the focusing NLS equation  arXiv:2002.11638 v1 

[l ,O,R] A.le Yacuanc, L.Oliver, J-C Raynal Journal of Math. Phys. 38 (1997) 3997-4015)

[M1] R.Minlos   On the point-like interaction between N fermions and another particle Moscow Math. Journal  11 (2011) 113-127

[M2] R.Minlos  A system of three quantum particles with point-like interactions Russian Math. Surveys  69b (2014) 539-564 

[T,S] K Ter-Martirosian , G Skorniakov  Three body problem for short range forces Sov. Phys. JETP  4 (1956) 648-661

\bigskip

Acknowlegments 

I benefitted much from correspondence with the late  R.Minlos; I want to thank  A.Michelangeli   for collaboration  at a very  early stage of this research.

\end{document}